\documentclass[12pt,a4paper]{article}
\usepackage[hscale=0.8,vscale=0.85]{geometry}
\usepackage{graphics}
\usepackage{stmaryrd}
\usepackage{amsfonts}
\usepackage{mathrsfs}
\begin{document}

\title{Thermodynamic equilibrium condition and the first law of thermodynamics for charged perfect fluids in electromagnetic and gravitational fields}

\author{
\footnotesize Kai Shi$^1$\thanks{Email: kaishi@mail.bnu.edu.cn},
Yu Tian$^2$\thanks{Email: ytian@ucas.ac.cn},
Xiaoning Wu$^3$\thanks{Email: wuxn@amss.ac.cn},
Hongbao Zhang$^1$\thanks{Email: hongbaozhang@bnu.edu.cn},
Chuanjia Zhu$^1$\thanks{Email: chuanjiazhu@mail.bnu.edu.cn}
\\
\footnotesize $^1$Department of Physics, Beijing Normal University, Beijing 100875, China\\
\footnotesize $^2$ School of Physics, University of Chinese Academy Sciences, Beijing 100049, China\\
\footnotesize $^3$ Institute of Mathematics, Chinese Academy of Sciences, Beijing 100190, China}

\date{}

\maketitle

\begin{abstract}
We provide a proof of the necessary and sufficient condition on the profile of the temperature, chemical potential, and angular velocity for a charged perfect fluid in dynamic equilibrium to be in thermodynamic equilibrium not only in fixed but also in dynamical electromagnetic and gravitational fields. In passing, we also present the corresponding expression for the first law of thermodynamics for such a charged star.
\end{abstract}

\section{Introduction}
In recent discussions of the generating functional for the equilibrium thermodynamic response of a relativistic fluid to the external $U(1)$ gauge field and gravitational field\cite{India,Meyer,Jensen,HK,KS1,KS2}, it is usually assumed that the local temperature satisfies Tolman's law, namely the redshifted local temperature is uniform throughout the whole fluid. The similar ansatz is also made over there for the profile of the local chemical potential. However, the validity of these assumptions has been proved only for the case in which the background spacetime is static. On the other hand, in support of the entropy principle for the self-gravitating charged perfect fluid\cite{Gao,FG1,FG2,YFJ},  only assuming Tolman's law, can one derive  the aforementioned ansatz for the local chemical potential from the equation of motion of the fluid. This may give one a misleading impression that these two assumptions are not independent of each other. To the best of our knowledge, Katz and Manor are the first to derive the uniformness of the redshifted temperature, chemical potential and angular velocity for the self-gravitating neutral perfect fluid on the same footing by the reasonable thermodynamic equilibrium requirement of the extremum of the total entropy at fixed total energy, total particle number, and total angular momentum\cite{KM}. Later on, such a derivation is much simplified in \cite{GSW} by resorting to the Iyer-Wald formalism developed in \cite{Wald1,IW,Iyer}.

Motivated by this, we intend to generalize such a derivation to the charged perfect fluid in a general stationary spacetime. Moreover, not only is our derivation made for the self-gravitating charged perfect fluid, but also for the charged perfect fluid in fixed electromagnetic and gravitational fields, corresponding to the aforementioned first case in which electromagnetic and gravitational fields are taken as external sources.

Below we shall follow the notations and conventions set in \cite{Wald} unless specified.

\section{Charged perfect fluids and dynamic equilibrium}
We start with the ordinary first law of thermodynamics
\begin{equation}
dE=TdS-pdV+\mu dN,\label{ofl}
\end{equation}
which together with the following extensive property
\begin{equation}
E(\lambda S, \lambda V, \lambda N)=\lambda E(S, V, N)
\end{equation}
gives rise to Euler's equation
\begin{equation}
E=TS-pV+\mu N.
\end{equation}
For our purpose, we would like to rewrite it and the first law of thermodynamics in a local way as follows
\begin{equation}
\rho+p=Ts+\mu n, \quad d\rho=Tds+\mu dn,
\end{equation}
whereby we further have
\begin{equation}
dp=sdT+nd\mu.
\end{equation}

Next let us consider a charged perfect fluid, by which we mean its energy momentum tensor and charge current take
the following form
\begin{equation}
T_{ab}=(\rho+p)u_au_b+pg_{ab}, \quad J_a=enu_a
\end{equation}
with $u^a$ the four velocity satisfying $u_au^a=-1$ and $e$ the constant charge carried by per particle under consideration. The dynamics of such a charged perfect fluid is governed by the following equations of motion
\begin{equation}
\nabla_aJ^a=0,\quad \nabla_aT^{ab}=F^{bc}J_c,
\end{equation}
with $F=dA$ the electromagnetic field strength. Here the first equation tells us the conservation of the total charge, which is equivalent to the conservation of the total particle number
\begin{equation}
N=\int_\Sigma nu^a\epsilon_{abcd}\equiv \int_\Sigma \mathbf{N}
\end{equation}
with $\Sigma$ any Cauchy surface and $\epsilon$ the volume element compatible with the metric. On the other hand, the contraction of $u_b$ with the second one gives us
\begin{equation}
0=-u_bF^{bc}J_c=u^a\nabla_a\rho+(\rho+p)\nabla_au^a=T\nabla_a(su^a)+\mu\nabla_a(nu^a)=T\nabla_a(su^a),
\end{equation}
which implies the conservation of the total entropy
\begin{equation}
S=\int_\Sigma su^a\epsilon_{abcd}\equiv\int_\Sigma\mathbf{S}.
\end{equation}

In what follows, we consider a general stationary, axisymmetric spacetime with the electromagnetic potential satisfying $\mathcal{L}_tA_a=\mathcal{L}_\varphi A_a=0$ for the timelike and axial Killing fields $t^a$ and $\varphi^a$, where our charged perfect fluid is in dynamical equilibrium if $\mathcal{L}_t u^a=\mathcal{L}_\varphi u^a=\mathcal{L}_t T=\mathcal{L}_\varphi T=\mathcal{L}_t\mu=\mathcal{L}_\varphi \mu=0$ and its four velocity is given by the following circular form
\begin{equation}
u^a=(t^a+\Omega\varphi^a)/|v|
\end{equation}
with $\Omega$ the angular velocity and
\begin{equation}
|v|^2=-g_{ab}(t^a+\Omega\varphi^a)(t^b+\Omega\varphi^b).
\end{equation}

With the above preparation, we shall investigate the necessary and sufficient condition on the profile of the temperature, chemical potential, and angular velocity for the thermodynamic equilibrium of such a charged perfect fluid in dynamic equilibrium. To this end, we first consider the case in which the background fields are non-dynamical, namely fixed.

\section{Fixed background fields}
Suppose that $\xi^a$ is either $t^a$ or $\varphi^a$, then we have
\begin{equation}
\nabla_a(T^{ab}\xi_b)=\xi_bF^{bc}J_c=(\xi\cdot dA)_c J^c=[\mathcal{L}_\xi A_c-d(\xi\cdot A)_c]J^c=-\nabla_a(J^aA^b\xi_b),
\end{equation}
where we have used the Cartan identity $\mathcal{L}_\chi\omega=\chi\cdot d\omega +d(\chi\cdot \omega)$ for any vector field $\chi^a$ and any form $\omega$ in the third step with the dot denoting the contraction of $\chi^a$ with the first index of the form. With this, we obtain the conservation of the total energy and total angular momentum of our charged perfect fluid as follows
\begin{equation}
E=-\int_\Sigma (T^{ae}+J^aA^e)t_e\epsilon_{abcd},\quad j=\int_\Sigma (T^{ae}+J^aA^e)\varphi_e\epsilon_{abcd}\equiv\int_\Sigma \mathbf{j}
\end{equation}
up to the constant shifts as $E\rightarrow E-ec_EN$ and $j\rightarrow j+ec_jN$ with $c_E=\mathcal{L}_t\lambda$ and $c_j=\mathcal{L}_\varphi\lambda$ under the residual gauge transformation $A_a+\nabla_a \lambda$ with $\nabla_a\mathcal{L}_t\lambda=\nabla_a\mathcal{L}_\varphi\lambda=0$. Because such shifts can be transferred to a constant shift of the chemical potential, corresponding essentially to the redefinition of the chemical potential, in what follows we shall not bother ourselves by such an irrelevant residual gauge dependence.

Note that our background fields are fixed, so the variation of the total energy is given by
\begin{eqnarray}
\delta E&=&-\int_\Sigma(\delta T^{ae}+\delta J^aA^e)(|v|u_e-\Omega\varphi_e)\epsilon_{abcd}\nonumber\\
&=&-\int_\Sigma|v|[-\delta\rho u^a-(\rho+p)\delta u^a+eA^eu_e\delta (nu^a)]\epsilon_{abcd}+\int_\Sigma \Omega(\delta T^{ae}+\delta J^aA^e)\varphi_e\epsilon_{abcd}\nonumber\\
&=&\int_\Sigma |v|T\delta (su^a)\epsilon_{abcd}+\int_\Sigma|v|(\mu-eA^eu_e)\delta(nu^a)\epsilon_{abcd}+\int_\Sigma \Omega(\delta T^{ae}+\delta J^aA^e)\varphi_e\epsilon_{abcd},\nonumber\\ \label{fl0}
\end{eqnarray}
where we have used $\delta u^au_a=0$ in the second step. Whence we obtain that the necessary and sufficient condition for our charged perfect fluid to be in thermodynamic equilibrium is given by
\begin{equation}
|v|T\equiv\tilde{T}=const.,\quad |v|(\mu-eA^eu_e)\equiv\tilde{\mu}=const., \quad \Omega=const.\label{ns}
\end{equation}
throughout the whole fluid.
Note that the thermodynamic equilibrium condition of our charged perfect fluid amounts to saying that the total entropy is required to be in the extremum at fixed total energy $E$, total angular momentum $j$, and total particle number $N$. Therefore it is obvious to see that Eq. (\ref{ns}) is the sufficient condition for our charged fluid to be in thermodynamic equilibrium. On the other hand, Eq. (\ref{ns}) also as the necessary condition for thermodynamic equilibrium can be argued by contradiction as follows. First, with our charged perfect fluid in thermodynamic equilibrium, one can always find a perturbation $\delta N_1=\delta j_1=0$ but $\delta S_1\neq0$ and $\delta E_1\neq0$. Now suppose that $\tilde{T}$ is not uniform throughout our charged fluid, then one can define its average as
\begin{equation}
\bar{\tilde{T}}=\frac{\int_\Gamma \tilde{T}\nu^a\epsilon_{abcd}}{\int_\Gamma \nu^a\epsilon_{abcd}}
\end{equation}
where $\Gamma$ is the subregion of $\Sigma$, through which our charged fluid permeates, and $\nu^a$ is the future directed timelike normal vector to $\Sigma$. Whence one can choose another perturbation
\begin{equation}
\delta\mathbf{S}_2|_\Gamma=\varepsilon(\tilde{T}-\bar{\tilde{T}})\nu^a\epsilon_{abcd}, \quad \delta\mathbf{S}_2|_{\Sigma/\Gamma}=0, \quad \delta\mathbf{N}_2|_\Sigma=0,\quad
\delta\mathbf{j}_2|_\Sigma=0
\end{equation}
with $\varepsilon$ a small parameter such that $\delta S_2=\delta N_2=\delta j_2=0$ and $\delta E_2=-\delta E_1$ with $\delta E_2$ given by Eq. (\ref{fl0}) as follows
\begin{equation}
\delta E_2=\varepsilon\int_\Gamma(\tilde{T}-\bar{\tilde{T}})^2\nu^a\epsilon_{abcd}.
\end{equation}
As a result, one can combine the above two perturbations such that $\delta E=\delta N=\delta j=0$ but $\delta S\neq0$, which contradicts with the fact that our charged fluid is in thermodynamic equilibrium. Therefore $\tilde{T}$ must be uniform throughout the fluid. By the same token, one can argue that $\tilde{\mu}$ and $\Omega$ must also be uniform. Whence the first law of thermodynamics for our charged perfect fluid can be expressed finally as
\begin{equation}
\delta E=\tilde{T}\delta S+\tilde{\mu}\delta N+\Omega\delta j. \label{fl}
\end{equation}
At first sight, the lack of the term like $p\delta V$ appears to be at odds with the ordinary first law of thermodynamics, saying Eq. (\ref{ofl}). However, on second thought, Eq. (\ref{fl}) is reasonable because the state parameters of our charged fluid are its energy, particle number, and angular momentum while the size of our charged fluid is totally determined by them through the natural boundary condition that the pressure becomes zero at its surface. Put it another way, the natural boundary condition $p=0$ at the fluid star surface provides a graceful exit for the $p\delta V$ term as it should be the case.

\section{Dynamical background fields}
For the case in which the background fields are not fixed but dynamical in the sense that they are affected by the backreaction of our charged perfect fluid through Einstein equation and Maxwell equation
\begin{equation}
H_{ab}\equiv G_{ab}-\frac{1}{2}T_{ab}^{EM}-\frac{1}{2}T_{ab}=0,\quad H^b\equiv \nabla_aF^{ab}+J^b=0,
\end{equation}
we would like to employ the Iyer-Wald formalism to address the thermodynamic equilibrium condition for our charged perfect fluid. As such, we consider the Lagrangian form
\begin{equation}
\mathbf{L}=(R-\frac{1}{4}F_{ab}F^{ab})\epsilon,
\end{equation}
whereby we have
\begin{equation}
\delta\mathbf{L}=E_{ab}\delta g^{ab}\epsilon+E^a\delta A_a\epsilon+d\mathbf{\Theta},
\end{equation}
where
\begin{equation}
E_{ab}=(G_{ab}-\frac{1}{2}T_{ab}^{EM})=G_{ab}-\frac{1}{2}(F_{ac}F_b{}^c-\frac{1}{4}F_{cd}F^{cd}g_{ab}),\quad E^a=\nabla_bF^{ba},
\end{equation}
and
\begin{equation}
\mathbf{\Theta}(\phi,\delta\phi)=\mathbf{\Theta}_{GR}+\mathbf{\Theta}_{EM}=w_{GR}\cdot\epsilon+w_{EM}\cdot\epsilon
\end{equation}
with $\phi$ denoting both the metric and electromagnetic potential and
\begin{equation}
w_{GR}^a=g^{ab}g^{cd}(\nabla_d\delta g_{bc}-\nabla_b\delta g_{cd}), w_{EM}^a=-F^{ab}\delta A_b.
\end{equation}
The Noether current is defined as follows
\begin{equation}
\mathbf{J}_\chi=\mathbf{\Theta}_{GR}(\mathcal{L}_\chi g)+\mathbf{\Theta}_{EM}(\mathcal{L}_\chi A)-\chi\cdot\mathbf{L}\label{def}
\end{equation}
with $\chi^a$ an arbitrary vector field. Whence one can show
\begin{equation}
\mathbf{J}_\chi=C_\chi\cdot\epsilon+d\mathbf{Q}_\chi \label{ide}
\end{equation}
with
\begin{equation}
C^a_\chi=(2E^a{}_b-E^aA_b)\chi^b, \quad \mathbf{Q}_\chi=\mathbf{Q}^{GR}_\chi+\mathbf{Q}^{EM}_\chi=-*d\chi-*FA_b\chi^b,
\end{equation}
where the star denotes the Hodge dual. By variation of both Eq. (\ref{def}) and Eq. (\ref{ide}) with $\chi^a$ fixed, one ends up with the following identity
\begin{equation}
d(\delta \mathbf{Q}_\chi-\chi\cdot\mathbf{\Theta})=\omega(\phi,\delta\phi,\mathcal{L}_\chi\phi)-\chi\cdot\epsilon( E_{ab}\delta g^{ab}+E^a\delta A_a)-\delta (C_\chi\cdot\epsilon)
\end{equation}
with
\begin{equation}
\omega(\phi,\delta_1\phi,\delta_2\phi)=\delta_1\mathbf{\Theta}(\phi,\delta_2\phi)-\delta_2\mathbf{\Theta}(\phi,\delta_1\phi).
\end{equation}
For our purpose, we shall choose $\chi^a$ to be $\xi^a$ in the previous section, namely a Killing vector field satisfying $\mathcal{L}_\xi A_a=0$. Accordingly, the above identity reduces to
\begin{equation}
d(\delta \mathbf{Q}_\xi-\xi\cdot\mathbf{\Theta})=-\xi\cdot\epsilon( E_{ab}\delta g^{ab}+E^a\delta A_a)-\delta (C_\xi\cdot\epsilon),
\end{equation}
which gives rise to the expression of the variation of the ADM mass and angular momentum as
\begin{eqnarray}
\delta M&\equiv&\delta\int_{S_\infty}(\mathbf{Q}_t-t\cdot\mathbf{B})=\int_{S_\infty}(\delta \mathbf{Q}_t-t\cdot\mathbf{\Theta})=\int_\Sigma [-t\cdot\epsilon( E_{ab}\delta g^{ab}+E^a\delta A_a)-\delta (C_t\cdot\epsilon)],\nonumber\\
\\
\delta j&\equiv& -\delta\int_{S_\infty}\mathbf{Q}_\varphi=\int_\Sigma \delta (C_\varphi\cdot\epsilon)=\int_\Sigma\varphi^b\delta[(T^a{}_b+J^aA_b)\epsilon_{acde}]\equiv\int_\Sigma \delta \mathbf{j},\label{amd}
\end{eqnarray}
where the sphere at infinity $S_\infty$ and $\Sigma$ are so chosen that $\varphi^a$ is tangent to them. Whence we further have
\begin{eqnarray}
\delta M&=&\int_\Sigma\{-|v|[( E_{ab}\delta g^{ab}+E^a\delta A_a)u^c\epsilon_{cdef}+u^b\delta( (2E^c{}_b-E^cA_b)\epsilon_{cdef})]+\Omega\delta\mathbf{j}\}.
\end{eqnarray}
Note that
\begin{eqnarray}
J^a\delta A_au^c\epsilon_{cdef}-u^b\delta(J^cA_b\epsilon_{cdef})&=&-eu^bA_b\delta \mathbf{N},\\
 -\frac{1}{2}T_{ab}\delta g^{ab}u^c\epsilon_{cdef}-u^b\delta (T^c{}_b\epsilon_{cdef})&=&\mu\delta \mathbf{N}+T\delta \mathbf{S},\label{lemma}
\end{eqnarray}
where the explicit derivation of Eq. (\ref{lemma}) can be found in \cite{GSW}. Therefore, we end up with
\begin{eqnarray}
\delta M&=&\int_\Sigma \{\tilde{T}\delta \mathbf{S}+\tilde{\mu}\delta\mathbf{N}+\Omega\delta\mathbf{j}-|v|[( H_{ab}\delta g^{ab}+H^a\delta A_a)u^c\epsilon_{cdef}+u^b\delta( (2H^c{}_b-H^cA_b)\epsilon_{cdef})]\}\nonumber\\
&=&\int_\Sigma (\tilde{T}\delta \mathbf{S}+\tilde{\mu}\delta\mathbf{N}+\Omega\delta\mathbf{j}).
\end{eqnarray}
Then following the same argument in the previous section, we obtain that the necessary and sufficient condition for the thermodynamic equilibrium of our charged fluid in dynamical background fields is also given by Eq. (\ref{ns}) and the corresponding first law of thermodynamics is given by Eq. (\ref{fl}).

However, here is a caveat. Different from the fixed background case, now the perturbations $\delta\mathbf{S}$, $\delta\mathbf{N}$ and $\delta \mathbf{j}$ are required to satisfy the linearized constraint equations. To justify our statement presented above, we are required to show they can actually be chosen in an arbitrary way. To this end, we first note that there exists a $t-\varphi$ reflection invariant Cauchy surface for our stationary background solution\cite{Carter}. To make our life simple, we shall choose $\Sigma$ to be such a Cauchy surface. Furthermore, we would like to fix once and for all our coordinate system in which the metric takes
\begin{equation}
ds^2=-\alpha^2d\tau^2+h_{ij}(dx^i+\beta^id\tau)(dx^j+\beta^jd\tau)
\end{equation}
with $\Sigma$ given by the surface of $\tau=0$.
As a result, the inverse of the metric is given by
\begin{equation}
g^{\mu\nu}=\left(
             \begin{array}{cc}
               -\frac{1}{\alpha^2} &  \frac{\beta^j}{\alpha^2}\\
               \frac{\beta^i}{\alpha^2} & h^{ij}-\frac{\beta^i\beta^j}{\alpha^2} \\
             \end{array}
           \right)
\end{equation}
with $h^{ij}$ the inverse of $h_{ij}$ and $\beta_k=h_{kj}\beta^j$, whereby we have
 \begin{equation}
 \nu_a=-\alpha(d\tau)_a, \quad \nu^a=\frac{1}{\alpha}[(\frac{\partial}{\partial \tau})^a-\beta^i(\frac{\partial}{\partial x^i})^a],
 \end{equation}
 and
 \begin{equation}
 h_{ab}=g_{ab}+\nu_a\nu_b=h_{ij}[(dx^i)_a+\beta^i(d\tau)_a][(dx^j)_b+\beta^j(d\tau)_b],\quad h^{ab}=g^{ab}+\nu^a\nu^b=h^{ij}(\frac{\partial}{\partial x^i})^a(\frac{\partial}{\partial x^j})^b.
 \end{equation}
 To facilitate our calculation, below we like to work with the gauge in which $\alpha=1$, $\beta^i=0$ and $\delta \alpha=\delta \beta^i=0$ on $\Sigma$. In addition, we shall work with the tensor density fields because the corresponding calculation turns out to be much simplified.

  Now by making use of the Gauss-Codazzi equation $2\nu_a\nu_bG^{ab}=R^{(3)}-K_{ab}K^{ab}+K^2$, $u^a=-u^b\nu_b\nu^a+\frac{u_b\varphi^b}{\varphi_c\varphi^c}\varphi^a$, Eq. (\ref{amd}), and Eq. (\ref{lemma}) as well as the background $K=0$ due to the fact that $K_{ab}$ is odd under the $t-\varphi$ reflection, we can express the linearized Hamiltonian constraint $\delta (2\sqrt{h}\nu_a\nu_bH^{ab})=0$ on $\Sigma$ as follows
 \begin{eqnarray}
 &&\hat{\epsilon}[-R^{(3)ij}(h)\delta h_{ij}+D^iD^j\delta h_{ij}-D^kD_k(h^{ij}\delta h_{ij})+h^{-1}\pi^{kl}\pi_{kl}h^{ij}\delta h_{ij}-2h^{-1}\pi_{ij}\delta \pi^{ij}\nonumber\\
 &&-2h^{-1}\pi^{ik}\pi^j{}_k\delta h_{ij}-\frac{1}{2}h^{-1}\pi^i\pi^j\delta h_{ij}+(D_kA^i-D^iA_k)D^{[k}A^{j]}\delta h_{ij}+\frac{1}{2}h^{-1}\pi^k\pi_kh^{ij}\delta h_{ij}\nonumber\\
 &&+E^{ab}\nu_a\nu_bh^{ij}\delta h_{ij}+\frac{1}{2}T^{ij}\delta h_{ij}-h^{-1}\pi_i\delta \pi^i-2D^{[i}A^{j]}D_{[i}\delta A_{j]}]-\frac{e\varphi_bu^b}{u_a\nu^a\varphi_c\varphi^c }\varphi^i\delta A_i\mathbf{N}\nonumber\\
 &&=-\frac{1}{u_a\nu^a}[(\mu-\frac{e\varphi_b\varphi^bA_d\varphi^d}{\varphi_c\varphi_c})\delta\mathbf{N}+T\delta\mathbf{S}+\frac{\varphi_bu^b}{\varphi_c\varphi^c}\delta \mathbf{j}].\label{c3}
 \end{eqnarray}
  Here $D$,  $\hat{\epsilon}=\sqrt{h}d^3x$ and $R^{(3)ij}$ are the derivative operator, the volume element and Ricci tensor associated with $h_{ij}$ on $\Sigma$ while $\pi^{ij}=\sqrt{h}(K^{ij}-Kh^{ij})$ and $\pi^i=\sqrt{h}F^{0i}$ with $K_{ij}$ the extrinsic curvature. In addition, $\mathbf{N}=\mathcal{N}d^3x$, $\mathbf{S}=\mathcal{S}d^3x$, and $\mathbf{j}=\mathcal{J}d^3x$.

Similarly, with the Codazzi-Maindardi equation $h_{ab}\nu_cG^{bc}=D_bK^b{}_a-D_aK$, one can express the linearized momentum constraints $\delta( 2\sqrt{h}h_{ab}\nu_cH^{bc})=0$ along the $\varphi^a$ direction and the direction perpendicular to $\varphi^a$ as
\begin{eqnarray}
&&\hat{\epsilon}\varphi^i[2D_k(h^{-\frac{1}{2}}\delta\pi^{kl}h_{li})+2h^{-\frac{1}{2}}\pi^{kl}D_k\delta h_{li}-h^{-\frac{1}{2}}\pi^{kl}D_i\delta h_{kl}+2D_k(h^{-\frac{1}{2}}\pi^{kl})\delta h_{li}\nonumber\\
&&+2h^{-\frac{1}{2}}\pi^jD_{[i}\delta A_{j]}+2D_{[i}A_{j]}(h^{-\frac{1}{2}}\delta \pi^j)]-e\mathbf{N}\varphi^i\delta A_i=eA_d\varphi^d\delta\mathbf{N}-\delta \mathbf{j},\label{c2}
\end{eqnarray}
and
\begin{eqnarray}
&&\hat{\epsilon}\varphi_\perp^i[2D_k(h^{-\frac{1}{2}}\delta\pi^{kl}h_{li})+2h^{-\frac{1}{2}}\pi^{kl}D_k\delta h_{li}-h^{-\frac{1}{2}}\pi^{kl}D_i\delta h_{kl}-2h^{lk}D_{[k}A_{j]}(h^{-\frac{1}{2}}\pi^j)\delta h_{li}\nonumber\\
&&+2h^{-\frac{1}{2}}\pi^jD_{[i}\delta A_{j]}+2D_{[i}A_{j]}(h^{-\frac{1}{2}}\delta \pi^j)]=\hat{\epsilon}\varphi^i_\perp h_{ij}\delta u^j(\rho+p)u_a\nu^a,\label{c4}
\end{eqnarray}
respectively, where $h_{ab}\nu_cH^{bc}=0$ is used Eq. (\ref{c4}).

On the other hand, we also have another linearized constraint equation $\delta (\sqrt{h}\nu_aH^a)=0$ from the $U(1)$ gauge field
\begin{equation}
\hat{\epsilon}D_i(h^{-\frac{1}{2}}\delta \pi^i)=e\delta\mathbf{N}.\label{c1}
\end{equation}
Next with the following form of the perturbation of the electromagnetic field and metric
\begin{equation}
\delta A_i=0,\quad \delta \pi^i=\sqrt{h}D^i\phi, \quad \delta h_{ij}=\psi h_{ij},\quad \delta \pi^{ij}=\sqrt{h}D^{(i}F^{j)}-\psi\pi^{ij},
\end{equation}
Eq. (\ref{c1}) reduces to
\begin{equation}
D_iD^i\phi=h^{-\frac{1}{2}}e\delta\mathcal{N},
\end{equation}
which has a unique solution that goes to zero at infinity given any prescribed perturbation $\delta \mathcal{N}$. Furthermore, Eq. (\ref{c2}) can be cast into
\begin{equation}
\varphi_i[D_kD^{(k}F^{i)}+D^{[i}A^{j]}D_j\phi]=\frac{h^{-\frac{1}{2}}}{2}(eA_d\varphi^d\delta\mathcal{N}-\delta \mathcal{J})\equiv f,
\end{equation}
which is automatically satisfied if we take
\begin{equation}
D_kD^{(k}F^{i)}=-D^{[i}A^{j]}D_j\phi+\frac{f\varphi^i}{\varphi_d\varphi^d}.
\end{equation}
There also exists a unique solution to the above equation that goes to zero at infinity. Similarly, the linearized Hamiltonian equation gives rise to
\begin{equation}
-D_iD^i\psi+\mathcal{M}\psi=h^{-\frac{1}{2}}\pi_{ij}D^{(i}F^{j)}+\frac{1}{2}h^{-\frac{1}{2}}\pi_iD^i\phi-\frac{1}{2u_a\nu^a}[(\mu-\frac{e\varphi_b\varphi^bA_d\varphi^d}{\varphi_c\varphi_c})\delta\mathcal{N}+T\delta\mathcal{S}+\frac{\varphi_bu^b}{\varphi_c\varphi^c}\delta \mathcal{J}]
\end{equation}
with
\begin{equation}
\mathcal{M}=h^{-1}\pi^{ij}\pi_{ij}+\frac{1}{4}h^{-1}\pi^i\pi_i+\frac{1}{2}D_{[i}A_{j]}D^{[i}A^{j]}+\frac{(u_a\varphi^a)^2}{2\varphi_c\varphi^c}(\rho+p)+\frac{1}{4}(\rho+3p),
\end{equation}
which is manifestly non-negative. So we also have a unique solution for $\psi$ which vanishes at infinity. Finally, Eq. (\ref{c4}) boils down into
\begin{equation}
-2\varphi^i_\perp[(D_j(h^{-\frac{1}{2}}\pi^j{}_i)+h^{-\frac{1}{2}}\pi^jD_{[i}A_{j]})\psi]=\varphi^i_\perp h_{ij}\delta u^j(\rho+p)u_a\nu^a,
\end{equation}
which is an algebraic equation for $\delta u^i_\perp$ and  can be readily fulfilled.

To summarize, with an arbitrary choice of $\delta\mathbf{N}$, $\delta \mathbf{S}$, and $\delta\mathbf{j}$, we have shown that the linearized constraint equations can always be satisfied, which validates our previous argument for the thermodynamic equilibrium condition of our charged fluid in the dynamical background fields.
\section{Conclusion}
By resorting to the extremum of the entropy at fixed energy, particle number and angular momentum, we have provided a proof of the necessary and sufficient condition on the profile of the temperature, chemical potential, and angular velocity throughout the charged perfect fluid for its thermodynamic equilibrium in both fixed and dynamical background fields. Accordingly, we also present the first law of thermodynamics for our charged fluid. Such a result can be further generalized in a obvious way to a variety of circumstances, which includes other dimensions, other higher derivative theories such as Gauss-Bonnet gravity theory and Dirac-Born-Infeld theory for electrodynamics, and other boundaries like finite boundary and AdS one.

\section*{Acknowledgements}
We are grateful to Bob Wald for his helpful discussions and valuable clarifications regarding his work. This work is partially supported by NSFC with Grant No.11575286, 11675015, 11731001,
11875095, 11975235, and 12075026.


\begin{thebibliography}{20}
\bibitem{India}N. Banerjee, J. Bhattacharya, S. Bhattacharyya, S. Jain, S. Minwalla, and T. Sharma, JHEP 09, 046(2012).
\bibitem{Meyer}K. Jensen, M. Kaminski, P. Kovtun, R. Meyer, A. Ritz, and A. Yarom, Phys. Rev. Lett. 109, 101601(2012).
\bibitem{Jensen} K. Jensen, R. Loganayagam, and A. Yarom, JHEP 05, 134(2014).
\bibitem{HK}J. Hernandez and P. Kovtun, JHEP 05, 001(2017).
\bibitem{KS1} P. Kovtun and A. Shukla, JHEP 10, 007(2018).
\bibitem{KS2}P. Kovtun and A. Shukla, Phys. Rev. D 101,  104051(2020).
\bibitem{Gao}S. Gao, Phys. Rev. D 84, 104023(2011).
\bibitem{FG1}X. Fang and S. Gao, Phys. Rev. D 90, 044013(2014).
\bibitem{FG2}X. Fang and S. Gao, Phys. Rev. D 92, 024044(2015).
\bibitem{YFJ} W. Yang, X. Fang, and J. Jing, Gen. Relativ. Gravit. 53, 81(2021).
\bibitem{KM}J. Katz and Y. Manor, Phys. Rev. D 12, 956(1975).
\bibitem{GSW}S. R. Green, J. S. Schiffrin, and R. M. Wald, Class. Quant. Grav. 31, 035023(2014).
\bibitem{Wald1}R. M. Wald, Phys. Rev. D 48, R3427(1993).
\bibitem{IW}V. Iyer and R. M. Wald, Phys. Rev. D 50, 846(1994).
\bibitem{Iyer}V. Iyer, Phys. Rev. D 55, 3411(1997).
\bibitem{Wald}R. M. Wald, {\it General Relativity}, The University of Chicago Press(Chicago, 1984).

\bibitem{Carter}B. Carter, J. Math. Phys. 10, 70(1969).


\end{thebibliography}
\end{document}